\documentclass[journal=jpcafh,manuscript=article]{achemso}
\usepackage{graphicx}
\usepackage{amsmath,amsfonts,amssymb}
\usepackage{gensymb}
\usepackage{caption}
\usepackage{subcaption}
\usepackage{xcolor} 
\usepackage[version=4]{mhchem}
\usepackage{booktabs}
\usepackage{braket}
\usepackage{layouts}
\SectionNumbersOn

\newpage
\title{Laser-Induced Electronic and Vibronic Dynamics in the Pyrene Molecule and its Cation}

\author{Katherine R. Herperger$^{\times , \circ}$}
\affiliation{University of Ottawa, Department of Physics, Ottawa ON K1N 6N5, Canada}
\alsoaffiliation{Humboldt-Universit\"at zu Berlin, Physics Department and IRIS Adlershof, 12489 Berlin, Germany}
\author{Jannis Krumland$^\times$}
\affiliation{Humboldt-Universit\"at zu Berlin, Physics Department and IRIS Adlershof, 12489 Berlin, Germany}
\author{Caterina Cocchi}
\affiliation{Humboldt-Universit\"at zu Berlin, Physics Department and IRIS Adlershof, 12489 Berlin, Germany}
\alsoaffiliation{Carl von Ossietzky Universit\"at Oldenburg, Institute of Physics, 26129 Oldenburg, Germany}
\email{caterina.cocchi@uni-oldenburg.de}

\date{\today}
\begin{document}

\maketitle
\makeatother{$^\times$ These authors contributed equally to this work.}

\makeatother{$^\circ$ Present Address: University of British Columbia, Department of Physics and Astronomy, Vancouver BC V6T 1Z1, Canada}

\newpage

\begin{abstract}
Among polycyclic aromatic hydrocarbons, pyrene is widely used as an optical probe thanks to peculiar ultraviolet absorption and infrared emission features. Interestingly, this molecule is also an abundant component of the interstellar medium, where it is detected via its unique spectral fingerprints. In this work, we present a comprehensive first-principles study on the electronic and vibrational response of pyrene and its cation to ultrafast, coherent pulses in resonance with their optically active excitations in the ultraviolet region. The analysis of molecular symmetries, electronic structure, and linear optical spectra is used to interpret transient absorption spectra and kinetic energy spectral densities computed for the systems excited by ultrashort laser fields. By disentangling the effects of the electronic and vibrational dynamics via \textit{ad hoc} simulations with stationary and moving ions, and, in specific cases, with the aid of auxiliary model systems, we rationalize that the nuclear motion is mainly harmonic in the neutral species, while strong anharmonic oscillations emerge in the cation, driven by electronic coherence. Our results provide additional insight into the ultrafast vibronic dynamics of pyrene and related compounds and set the stage for future investigations on more complex carbon-conjugated molecules.

\begin{figure}[H]
\caption*{\textbf{TOC Graphic}}
    \centering
    \includegraphics{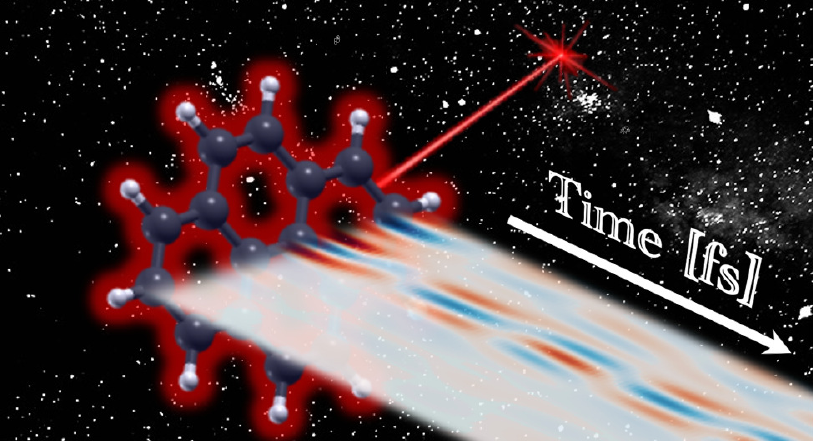}
\end{figure}
 
\end{abstract}

\newpage

\section{Introduction}

Polycyclic aromatic hydrocarbons (PAHs) are a prominent class of molecules that is relevant in diversified scientific areas, ranging from astrochemistry~\cite{salama1992pyrene,tielens2008interstellar,gredel2011abundances,peeters2021spectroscopic} to opto-electronics~\cite{wu2007,yuan2018engineering,anwar2019recent,yuan2020bright}, up to photocatalysis~\cite{tsai2020nitrogen}.
The chemical versatility of these systems is largely responsible for their success~\cite{Dias1985,drummer2021}: their electronic and optical properties can be flexibly modulated by substitution~\cite{dou2012,farrell2019,chen2019homo} and functionalization~\cite{naghavi2011theoretical,cocchi2011designing,cocchi2011optical,kaipio2012,cocchi2012,yuan2020bright}, and efficient nanostructures for molecular electronics can be obtained by vertical assembly of these moieties~\cite{feng2009towards,feng2009,pisula2010}.
The extended distribution of $\pi$ electrons is another peculiar feature of PAHs which crucially determines their electronic structure and optical activity~\cite{stein1987pi,philpott2009geometry,Fowler2007} also beyond the linear regime~\cite{zeng2011,papadakis2018}.

Among the experimental techniques commonly adopted to fingerprint PAHs, vibrational spectroscopies are worth a special mention. 
For example, resonance Raman spectroscopy is particularly powerful to detect the signatures of individual molecules, in agreement with corresponding simulations.~\cite{jens-scha06jpca,neug+05jpca}
Broadly speaking, carbon-based materials are in fact characterized by distinct vibrational modes~\cite{clar1941quantenmechanische} which are highly sensitive to the slightest changes in the structural conformation of the molecules as well as in their electronic distribution~\cite{pimenta2007studying,zhang2015combined,Dontot2020}.
Vibrational signatures are not only largely employed to recognize graphitic molecules and nanostructures~\cite{li2008dirac,ferrari2013raman,bokobza2014raman}; they are also used to probe the chemical composition of the interstellar medium~\cite{leger1989physics,zhang2014viability}.

When laser-driven spectroscopies emerged, PAHs were among the first systems to be systematically investigated with these new techniques~\cite{fleming1986chemical}, disclosing the evolution and the lifetime of their excited states~\cite{Foggi1995,baba2009vibrational}.
Recent advances in the field of ultrafast spectroscopies have finally enabled probing the charge carrier dynamics of these systems on their natural, femtosecond time scale~\cite{rozz+13natcom,falk+14sci,de2016tracking,borrego2018ultraviolet,borrego2019two,li2021ultrahigh}.
In this way, it is possible not only to monitor the ``birth" of the electronic excitations but also to investigate the coupling between electronic and vibrational degrees of freedom with unprecedented resolution. 
In the last decade, these experimental advances have been accompanied by the development of first-principles methods that are able to accurately reproduce and interpret these phenomena atomistically.
In particular, time-dependent density-functional theory (TDDFT)~\cite{Runge1984} in its real-time implementation and in conjunction with the Ehrenfest molecular dynamics scheme~\cite{marques2004,rozzi2017} is currently considered one of the most reliable and versatile approaches for describing the electron-nuclear dynamics of molecules and materials in the sub-picosecond timescale~\cite{otobe2009first,de2013simulating,PhysRevLett.113.087401,PhysRevB.90.174303,zhang2017manipulation,sato2018ab,liu2020manipulation,jacobs2020ultrafast,Krumland2020}. 

In this work, we investigate from first principles the ultrafast electronic and vibrational dynamics of pyrene (\ce{C16H10}), a widely studied member of the PAH family, in its neutral form as well as in its positively charged configuration, which is known for its enhanced electronic and vibrational activity~\cite{kira1971pyrene,kim2001single}.
Pyrene is characterized by intense absorption and emission bands in the near ultraviolet (UV) region, and by high and long-lived fluorescence yield~\cite{dawson1968,kropp1969radiative}.
The peculiar photo-response of this molecule makes it \textit{per se} an interesting compound for organic electronics (see, \textit{e.g.}, Ref.~\citenum{figueira2011pyrene} for review) as well as an efficient fluorescent probe~\cite{valdes1990pyrene,matsui1999characterization,nagatoishi2005pyrene}.
After the analysis of the symmetries of the system, its electronic structure, and linear optical spectra, we inspect the vibrational response upon resonant laser excitations, identifying the modes that are most prominently stimulated by the pulse. 
We clarify that the nuclear motion is mainly harmonic in the neutral species, while anharmonicities emerge in the cation.
The latter effects are a result of electronic forces arising from quantum-state interferences.
From the transient absorption spectra computed by selectively enabling and disabling the nuclear motion, we identify the contribution of the vibrational degrees of freedom to the ultrafast dynamics of the systems.

\section{Methodology} \label{sec:methods}

\subsection{Theoretical Background}

The results obtained in this work are based on real-time time-dependent density-functional theory (RT-TDDFT). We employ spin-restricted and spin-unrestricted formalism for neutral and cationic pyrene, respectively. The time-dependent (TD) electron density,
\begin{equation}
    \rho(\mathbf{r},t) = \rho_\uparrow(\textbf{r},t)+\rho_\downarrow(\textbf{r},t) = \sum_{\sigma =\uparrow,\downarrow}  \sum^{N_{\sigma}}_{i} \left | \psi_{i\sigma}(\mathbf{r},t) \right |^2,
\end{equation}
where $\sigma$ is the spin index, is calculated from the $N = N_\uparrow + N_\downarrow$ occupied Kohn-Sham (KS) orbitals, $\psi_{i\sigma}(\mathbf{r},t)$, which are propagated using the TD Kohn-Sham equation~\cite{Runge1984},
\begin{equation}
    i \frac{\partial}{\partial t} \psi_{i\sigma}(\mathbf{r},t) = \left ( - \frac{\nabla^2}{2} + v_{\mathrm{KS},\sigma}[\rho_\uparrow,\rho_\downarrow](\mathbf{r},t)\right ) \psi_{i\sigma}(\mathbf{r},t).
\end{equation}
The effective potential, 
\begin{equation}\label{eqn:ks}
    v_{\mathrm{KS},\sigma}[\rho_{\uparrow},\rho_{\downarrow}](\mathbf{r},t) =  v_{\mathrm{en}}(\mathbf{r},t) + v_{\mathrm{ext}}(\mathbf{r},t) + \int\text d^3 r' \frac{\rho(\mathbf{r'},t)}{\left | \mathbf{r} - \mathbf{r'} \right |} + v_{\mathrm{xc},\sigma}[\rho_{\uparrow},\rho_{\downarrow}](\mathbf{r},t),
\end{equation}
is composed of four terms: the electron-nuclear interaction potential ($v_{\mathrm{en}}$), the external potential ($v_{\mathrm{ext}}$), the Hartree potential, and the exchange-correlation potential ($v_{\mathrm{xc},\sigma}$). For the latter, the adiabatic local spin density approximation (ALSDA)~\cite{alda1,alda2,alda3,alda4} is adopted. In the length gauge, the external potential becomes
\begin{equation}
    v_{\mathrm{ext}}(\mathbf{r},t) = \mathbf{r} \cdot \mathbf{E}(t),
\end{equation} 
where $\mathbf{E}(t)$ is the incident TD electric field. To calculate linear absorption spectra, electron dynamics are initiated at $t=0$~fs by a broadband probe pulse
\begin{equation}
    \mathbf{E}(t) = \mathbf{E}_{\mathrm{probe}}(t) = \hat{\mathbf{n}} \, \kappa \, \delta (t),
\end{equation}
given by a Dirac delta function $\delta (t)$ of strength $\kappa$ and polarization $\hat{\mathbf{n}} \in \{\hat{\mathbf{x}},\hat{\mathbf{y}},\hat{\mathbf{z}}\}$. At each time step, the induced dipole moment is calculated as
\begin{equation}
    \mathbf{d}(t) = - \int \text d^3 r\,\mathbf{r} \left [ \rho (\mathbf{r},t)  - \rho_{\mathrm{GS}} (\mathbf{r}) \right ].
    \label{eqn:dens}
\end{equation}
In Eq.~\eqref{eqn:dens}, the integrand is the product of the position vector $\mathbf{r}$ and the difference between the TD and time-independent ground-state (GS) electron densities. The linear absorption spectrum is proportional to the imaginary component of the Fourier transform of $\textbf{d}(t)$, if a sufficiently small value of $\kappa$ is set.\cite{yabana} Large values of $\kappa$ are known to lead to optical nonlinearities,\cite{cocchi2014,Guandalini2021} but this aspect will not be considered in the present work.

In order to analyze the electronic dynamics associated with a particular excited state, the molecule is excited with a 
Gaussian-enveloped pump pulse, 
\begin{align}
    \textbf{E}(t) = \textbf{E}_{\text{pump}}(t) = \hat{\mathbf{n}}E_0e^{-(t-t_\mu)^2/2t_\sigma^2}\cos(\omega_pt),
\end{align}
where $E_0$ is the peak amplitude, $t_\mu$ and $t_\sigma$ are the mean time and standard deviation of the Gaussian envelope function, respectively, $\hat{\mathbf{n}} \in \{\hat{\mathbf{x}},\hat{\mathbf{y}},\hat{\mathbf{z}}\}$ is the polarization direction, and $\omega_p$ is the carrier frequency. $\hat{\mathbf{n}}$ and $\omega_p$ are chosen to ensure resonance with specific excitations in the linear-response spectrum. The pump pulse transfers the system into a non-stationary superposition between the GS and the targeted excited state, although the finite bandwidth associated with the finite pulse can lead to the population of energetically close-by states as well. 
The electron dynamics for fixed nuclei are analyzed by considering the time-dependent populations
\begin{equation}\label{eqn:orbitalProjection}
 P_{i\sigma}(t) = \sum_j^{N_\sigma}\left|\int\text d^3r\,\psi_{i\sigma}^*(\mathbf{r})\psi_{j\sigma}(\mathbf{r},t)\right|^2
\end{equation}
 of the occupied and virtual GS orbitals $\psi_{i\sigma}(\mathbf{r})$, which are the eigenstates of the KS Hamiltonian.

To calculate transient absorption spectra, the two aforementioned approaches~\cite{Krumland2020} are combined. With both pump and probe, the total electric field becomes
\begin{equation}
    \mathbf{E}(t) = \mathbf{E}_{\mathrm{pump}}(t) + \mathbf{E}_{\mathrm{probe}}(t-t_0),
\end{equation}
where $t_0$ is the time delay. We use phase-cycling\cite{hamm2011,seidner} to isolate the absorptive third-order contributions to the dipole moment induced by the pump and the probe. In this approach, the same calculation is performed twice: in the second run, a constant phase of $\pi$/2 is added to the pump field. The resulting dipole moments are averaged, and pump-only contributions are subtracted. A sine transformation of this quantity yields the non-equilibrium absorption cross section as a function of energy. Since all fields are weak, the changes occurring in the spectrum are small, and a reference spectrum is subtracted to obtain the differential cross section, $\Sigma=\Sigma(\omega, t_0)$. This reference spectrum is either the one at $t_0=0$~fs (linear absorption spectrum), or the one at $t_0=20$~fs (post-pulse non-equilibrium spectrum).

RT-TDDFT results are complemented by linear-response time-dependent density-functional theory (LR-TDDFT) calculations solving the Casida equation\cite{casida2012, casida1996}:
\begin{gather}
    \begin{pmatrix}
    \mathbf{A} & \mathbf{B} \\ 
    \mathbf{B^*} & \mathbf{A^*}
    \end{pmatrix}
    \begin{pmatrix}
    \vec{X}\\ 
    \vec{Y}
    \end{pmatrix}
    = 
    \omega
    \begin{pmatrix}
    \mathbf{1} & \mathbf{0} \\ 
    \mathbf{0} & \mathbf{-1}
    \end{pmatrix} 
    \begin{pmatrix}
    \vec{X}\\ 
    \vec{Y}
    \end{pmatrix}.   
    \label{eqn:casida}
\end{gather}
For transitions between orbitals $\psi_i(\mathbf{r})$, excitation and de-excitation coefficients are given by $\vec{X}$ and $\vec{Y}$, respectively. The elements of matrices $\mathbf{A}$ and $\mathbf{B}$ are\cite{Ullrich2011}
\begin{equation}
    A_{im\sigma_1,jn\sigma_2} = \delta_{i,j} \delta_{m,n}\delta_{\sigma_1,\sigma_2} (\epsilon_m - \epsilon_i) +  \bra{i\sigma_1,j\sigma_1}f_{\mathrm{Hxc}}^{\sigma_1, 
\sigma_2}\ket{m\sigma_2,n\sigma_2}
\end{equation}
and
\begin{equation}
    B_{im\sigma_1,jn\sigma_2} = \bra{i\sigma_1,n\sigma_2}f_{\mathrm{Hxc}}^{\sigma_1, \sigma_2}\ket{m\sigma_2,j\sigma_1},
\end{equation}
respectively, where $\epsilon_k$ are the KS eigenvalues, $\sigma_1$ and $\sigma_2$ the spin indices, and 
\begin{equation}
    f_{\mathrm{Hxc}}^{\sigma_1, 
\sigma_2}(\mathbf{r},\mathbf{r'}) = \frac{1}{\left | \mathbf{r} - \mathbf{r'} \right |} + \frac{\delta \nu_{\mathrm{xc},\sigma_1}[\rho](\mathbf{r})}{\delta \rho_{\sigma_2}(\mathbf{r'})} \bigg\rvert_{\rho = \rho_g}
\end{equation}
the Hartree-exchange-correlation kernel.
The solution of Eq.~\eqref{eqn:casida} yields the excitation energy eigenvalues $\omega$ and the eigenvectors $\vec{X}$ and $\vec{Y}$, which contain information about the composition of each excitation.

Molecular dynamics simulations are performed using the Ehrenfest scheme in conjunction with RT-TDDFT\cite{marques2004,marques2012,Andrade2009}. In this mean-field approach, nuclei are propagated classically according to the equation of motion
\begin{equation}\label{eq.ehrenfest_eom}
    M_J \frac{d^2}{dt^2} \mathbf{R}_J = -\nabla_{\mathbf{R}_J}[v_{\mathrm{ext}}(\mathbf{R}_J,t) + v_\text{nn}(R(t)) + \int \text d^3 r \rho(\mathbf{r},t) v_{\mathrm{en}}(\mathbf{r},R(t))],
\end{equation}
where $M_J$ and $\mathbf{R}_J$ are the mass and position vector of nucleus $J$, respectively, and $R(t)$ stands for the set of all $\mathbf{R}_J$; $v_{\mathrm{ext}}(\mathbf{R}_J,t)$ is the external potential, $v_\text{nn}(R(t))$ the electrostatic energy among the nuclei, and $v_\text{en}(\textbf{r},R(t))$ the electrostatic energy between the nuclei and an electron at position $\textbf{r}$. Electrons are treated quantum-mechanically according to the RT-TDDFT approach described above. Both the nuclear and electronic subsystems evolve in real time.

To analyze the nuclear motion, we determine the normal modes of vibration for the GS minimum configuration, $R^0=\{\textbf{R}^0_I\}$. This corresponds to expanding the internal potential terms with $\rho=\rho_\text{GS}$ in Eq.~\eqref{eq.ehrenfest_eom} around $R^0$, up to the second order for displacements $\Delta R=\{\Delta \textbf{R}_I\}$:
\begin{align}
    v_\text{nn}(R^0+\Delta R) + \int\text d^3 r \rho_{\mathrm{GS}}(\mathbf{r};R^0+\Delta R) &v_{\mathrm{en}}(\mathbf{r},R^0+\Delta R) \nonumber\\&\approx v_0 + \frac 12\sum_{I\mu,J\nu}k_{I\mu,J\nu}\Delta R_{I\mu}\Delta R_{J\nu},
\end{align}
where $v_0$ is the constant zeroth-order term. The Hessian computed at $R^0$,
\begin{equation}\label{eqn:spring}
    k_{I\mu,J\nu} = \left.\frac{\partial^2}{\partial R_{I\mu}\partial R_{J\nu}}\left[v_\text{nn}(R) + \int\text d^3 r \rho_{\mathrm{GS}}(\mathbf{r};R) v_{\mathrm{en}}(\mathbf{r},R)\right]\right|_{R=R^0},
\end{equation}
depends on the first parametric derivative of the electron density with respect to the nuclear coordinates, and can be calculated via density functional perturbation theory. Diagonalization of the mass-weighted Hessian, $k_{I\mu,J\nu}/\sqrt{M_IM_J}$, yields the matrix $T_{k,J\mu}$ that transforms mass-weighted Cartesian nuclear displacements and velocities into the normal mode basis:
\begin{subequations}\label{eqn:normalTrafo}
\begin{align}
    {\cal Q}_k &= \sum_{J\nu}T_{k,J\nu}\sqrt{M_J}\Delta R_{J\nu}   \\
    \dot{\cal Q}_k &= \sum_{J\nu}T_{k,J\nu}\sqrt{M_J}\dot{R}_{J\nu}.
\end{align}
\end{subequations}

Furthermore, this diagonalization provides us with the corresponding fundamental frequencies of vibration, $\{\nu_k^0\}$. Given a nuclear trajectory $R=R(t)$, we calculate the mode-resolved kinetic energy spectral density (KESD) as 
\begin{equation}\label{eqn:power}
    S_k(\nu) = \left|\int\text dt\,e^{-2\pi i\nu t}\dot{\cal Q}_k(t)\right|^2,
\end{equation}
which is the Fourier transform of the normal velocity autocorrelation function. Considering this quantity in relation with the fundamental frequency $\{\nu_k^0\}$ of the corresponding vibration allows us to pinpoint anharmonic effects. The total KESD is given by $S(\nu) = \sum_kS_k(\nu)$.

We employ the basis of GS normal modes also for the investigation of dynamics involving excited states. Such projections contain information about corresponding Duschinsky rotations and frequency shifts, which have been important in other theoretical studies of pyrene\cite{egid+14jctc,avil+13jctc}. However, we apply laser pulses causing excited-state populations of only $\sim$1-2\%, which means -- as a consequence of the mean-field nature of Ehrenfest dynamics -- that the nuclear dynamics are determined mainly by the GS normal vectors and frequencies. Inspecting $S_k(\nu)$ in terms of absolute values therefore reveals little about the difference between normal modes in the ground and excited states. Distinguishing these differences would require a reference simulation entirely on the ground-state potential energy surface. Due to the rigidity of the pyrene molecule, it is unlikely that Duschinsky rotations and frequency shifts play a significant role in our case, therefore we do not follow this path.

\subsection{Computational Details}
\label{ssec:comput}

All \textit{ab initio} calculations are performed with the OCTOPUS code~\cite{oct1,oct2,oct3,oct4}, unless otherwise specified. Herein, the simulation box is defined by interconnecting spheres of radius 5~\r{A} for each atom, which encloses a real-space regular cubic mesh with a spacing of 0.2~\r{A}. First-principles norm-conserving Troullier-Martins pseudopotentials are used.\cite{troullier1991}.
In the TD runs, the propagator is based on the approximated enforced time-reversal symmetry scheme~\cite{castro2004}. A time step of approximately 2.0~as is sufficiently small to describe the electron dynamics. 

The ground state geometries for neutral and cationic pyrene (see Tables~S1 and S2 in the Supporting Information, SI) are optimized iteratively, until interatomic forces are below the threshold of $10^{-5}$~eV/\r{A}. The discrete excited state energies are obtained from LR-TDDFT calculations. Linear absorption spectra computed from RT-TDDFT comprise a propagation time of 15 fs following the kick at $t=0$~fs, including in the post-processing a phenomenological damping in the form of a third-order polynomial.
For the calculation of transient absorption spectra (TAS), the pump field is described by a linearly-polarized Gaussian envelope function peaked at $t_\mu=$~12~fs with a standard deviation of $t_\sigma=$~3~fs and a peak amplitude $E_0$ corresponding to an intensity of 10$^{10}$ W/cm$^2$. The carrier frequency of the pump is determined by the excitation energies from the linear absorption spectra. Every 1~fs, the system is probed by a  broadband, instantaneous ``kick''. After each kick, the system is further propagated for 10~fs. In TAS calculations coupled with the Ehrenfest nuclear dynamics, nuclei are initially at rest. 

The code Gaussian~16\cite{g16} is adopted in two instances, namely to optimize excited-state geometries and to compute the normal modes of the pyrene cation, checking that the normal mode displacements in the neutral molecule and the cation are similar.
In these calculations, the ALSDA is employed together with a 6-311++G(d,p) basis set. 
The package XCrySDen\cite{xcrysden} is used to visualize structures and molecular orbitals. In the latter case, blue and red isosurfaces are representative of negative and positive regions.
The isovalues used were either $\pm$0.03~\AA$^{-3/2}$ or $\pm$0.035~\AA$^{-3/2}$, depending on the grid value range.

\section{Results and Discussion} \label{sec:results}

\subsection{Ground-State Characterization}

\begin{figure}[]
    \centering
    \includegraphics[width=0.5\textwidth]{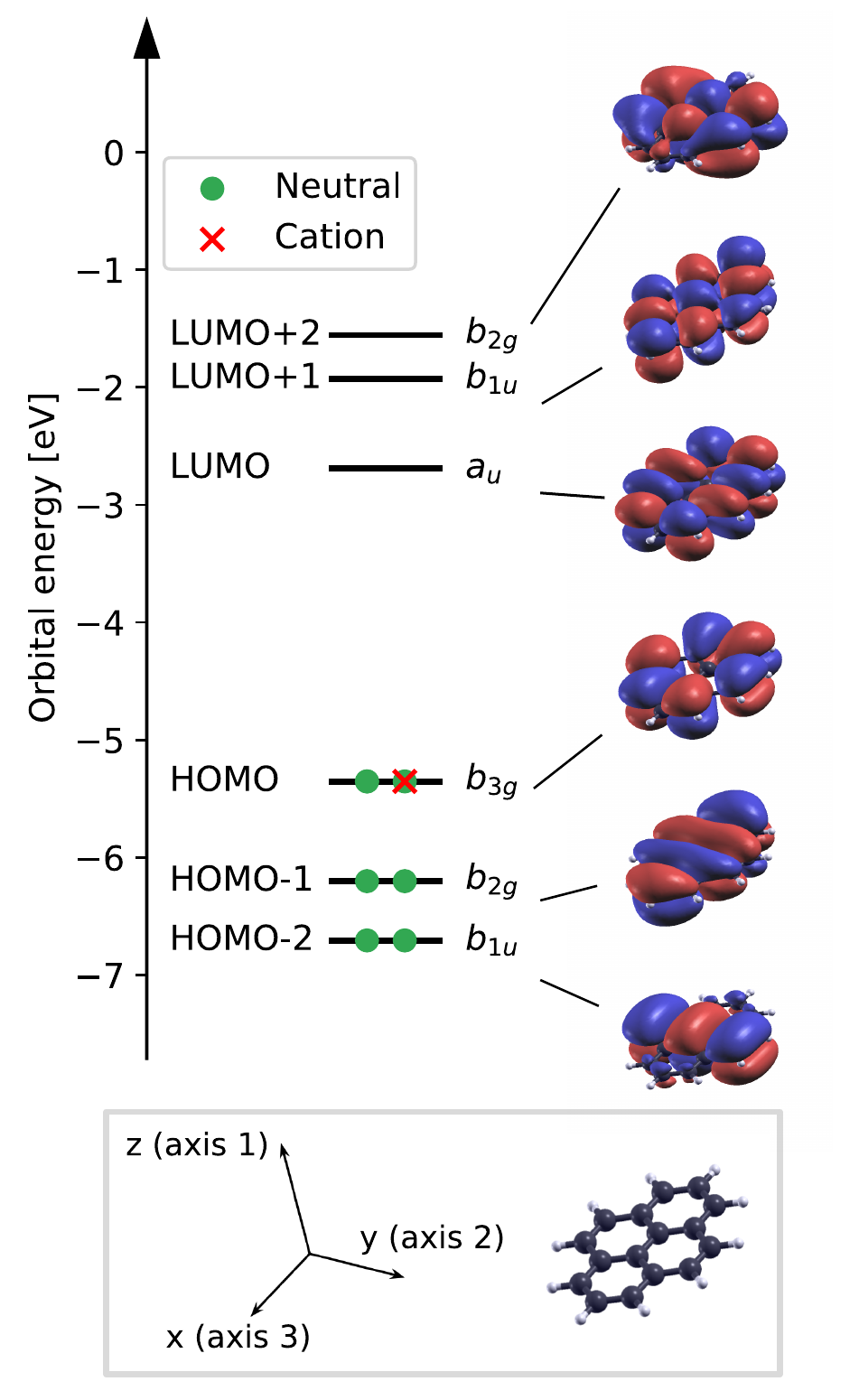}
    \caption{Top: Energy levels (left) and molecular orbitals (right) of neutral pyrene. The electron occupation in the ground state is marked by green circles. The red cross denotes the electron removed in the cation. Bottom: Geometry of pyrene in the adopted coordinate system with C atoms in dark grey and H atoms in white.}
    \label{fig:nrg}
\end{figure}

The optimized geometries of neutral and cationic pyrene have $D_{2h}$ symmetry, meaning that they are invariant with respect to the following symmetry elements: identity ($E$), three 2-fold axes of rotation [$C_2$(x),$C_2$(y),$C_2$(z)], inversion symmetry $i$, and horizontal mirror planes ($\sigma_h$(xy),$\sigma_h$(xz),$\sigma_h$(yz)).
The occupied and unoccupied molecular orbitals of neutral pyrene in the vicinity of the gap have $\pi$ and $\pi^*$ character, respectively (see Figure \ref{fig:nrg}), as expected for polycyclic aromatic hydrocarbons~\cite{BOSCHI1972,casanova2013,cocchi2013,cocchi2014jpca}. The cation, which has one less electron in the HOMO, is therefore characterized by singly occupied and unoccupied molecular orbitals at the frontier (SOMO and SUMO, respectively). Energetically, the orbitals in the cation are downshifted by approximately 4~eV with respect to those of the neutral species, although they preserve the same spatial distribution.
We describe the orbital symmetry denoting, as usual, $a$ as the rotation symmetry with respect to all three axes, and $b_n$ as the rotation symmetry only with respect to the axis $n$ (see Figure \ref{fig:nrg}). Additionally, all orbitals can be distinguished by parity due to the presence of inversion symmetry: \textit{gerade} orbitals are left unchanged by the inversion operation, while \textit{ungerade} orbitals undergo a change of sign upon inversion. 

For comparison with experimental references, we compute the ionization potential (IP) as the difference between the total energy obtained for the neutral molecule and its cation.
For the vertical IP, obtained by considering the total energy of the cation in the geometry of the neutral species, we get 7.44~eV, which matches well with the experimental value of 7.41~eV~\cite{Clar1976,BOSCHI1972}.
For the adiabatic IP, where the energy of the cation corresponds to the one in its relaxed geometry, we obtain 7.40~eV, in excellent agreement with the literature: 7.4064 $\pm$ 0.0007~eV was measured by Zhang \textit{et al.}~\cite{Zhang2010}, and 7.415~$\pm$~0.01 eV by Mayer and coworkers~\cite{mayer2011}.

\subsection{Linear absorption spectra}

\begin{figure}[]
    \centering
    \includegraphics{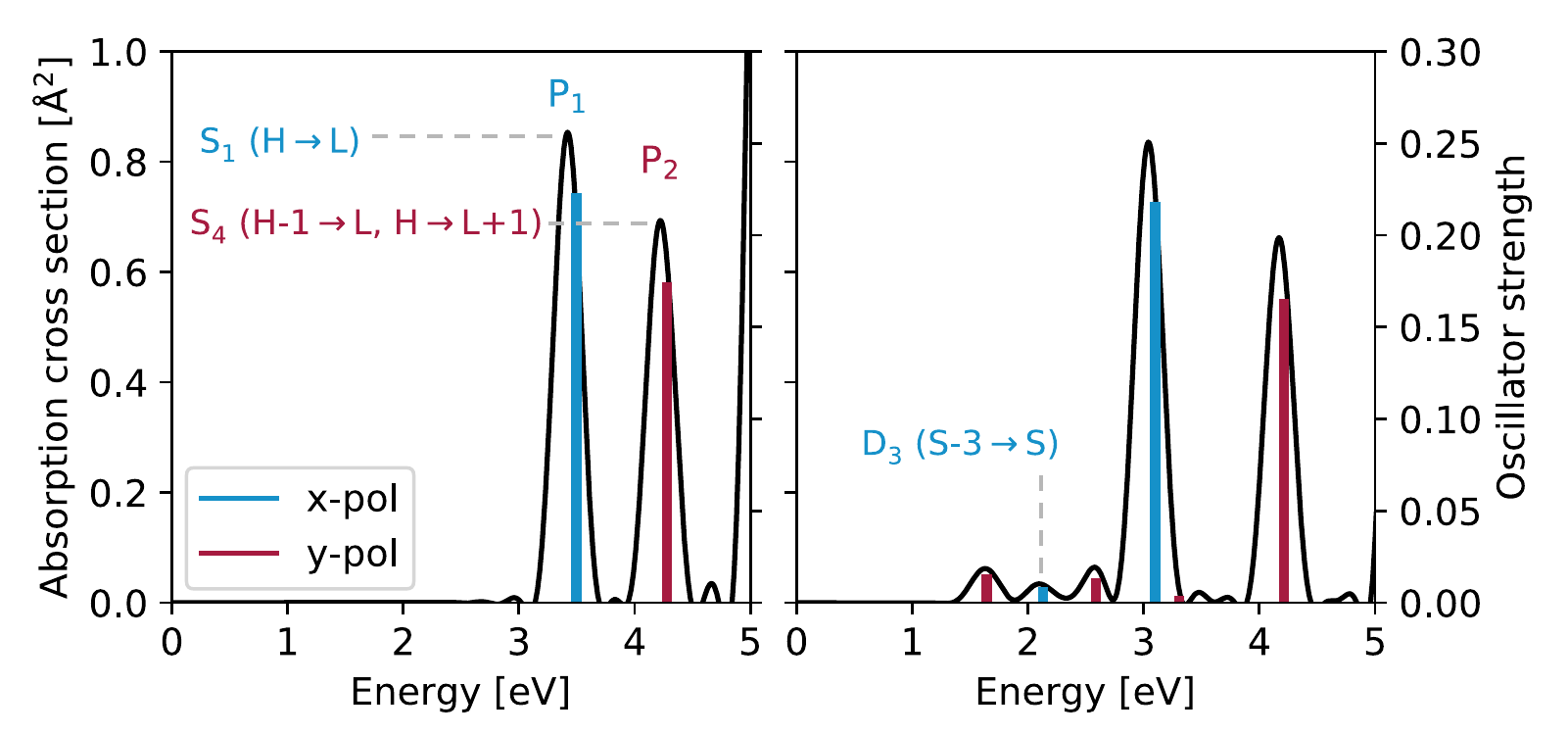}
    \caption{Linear absorption spectra computed from RT-TDDFT for neutral pyrene (left) and cationic pyrene (right). LR-TDDFT results are displayed as vertical bars coloured according to the respective polarization and with height indicative of their oscillator strength (scale bar on the right). The composition of selected excitations in terms of single-particle orbitals (H stands for HOMO, L for LUMO, and S for SOMO) is indicated in the spectra.}
    \label{fig:abs_spec}
\end{figure}

We begin the analysis of the excited-state properties of pyrene and its cation by examining their linear absorption spectra (see Figure~\ref{fig:abs_spec}). 
The spectrum of the neutral molecule (left) is dominated by two intense resonances. The lower-energy peak at 3.42~eV is labeled \ce{P1}, and corresponds to a singlet excitation polarized along $\hat{x}$, \textit{i.e.} the long axis of the molecule (see Figure~\ref{fig:nrg}, bottom). Our LR-TDDFT results reveal that it arises mainly from the transition from the HOMO to the LUMO (details reported in the SI, Table~S3).
The second peak, labeled \ce{P2}, corresponds to the fourth excited state \ce{S4}, and appears at 4.22~eV. It is $\hat{y}$-polarized and arises from the \textit{constructive} superposition of the HOMO$-$1~$\rightarrow$~LUMO and HOMO~$\rightarrow$~LUMO+1 transitions (see Table~S3 in the SI). 

The experimentally observed ordering of the energetically close excited states \ce{S1} and \ce{S2}~\cite{tanaka1965bcsj,baba2009vibrational} is incorrectly predicted by many theoretical methods,\cite{dierksenGrimme2004jcp,bito+2000cpl,Zhang2010} including ALSDA. 
Within density functional theory, the ordering is directly affected by the choice of the exchange-correlation functional. While our results denote the 1$^1$B$_{\mathrm{3u}}$ and 1$^1$B$_{\mathrm{2u}}$ states as \ce{S1} and \ce{S2}, respectively, the correct order observed experimentally (\ce{S1}=1$^1$B$_{\mathrm{2u}}$ and \ce{S2}=1$^1$B$_{\mathrm{3u}}$) can be reproduced using a range-separated hybrid functional such as CAM-B3LYP\cite{Crawford2011,YANAI200451}. 
This given, we do not expect the wrong ordering to be problematic, since vibrationally mediated transitions between states of B$_\text{3u}$ and B$_\text{2u}$ symmetry cannot occur in our simulations:
The rate of such transitions is determined by the matrix element $\langle \text B_\text{3u}|\hat V({\cal Q})|\text B_\text{2u}\rangle$, which is non-zero only if the nuclear displacements $\cal Q$ associated with the vibration-induced perturbation $\hat V({\cal Q})$ are $\text B_\text{1g}$ symmetric. 
However, as we will see in the following, modes of such symmetry do not participate in the nuclear dynamics. From a quantum-mechanical perspective, modes with $\text B_\text{1g}$ symmetry could come into play through the presence of zero-point energy, which would imply fluctuations of $\text B_\text{1g}$-symmetric nuclear positions and momenta around the classical minimum at ${\cal Q}=\dot{\cal Q}=0$.
However, such vibronic effects are not captured in the single-trajectory Ehrenfest scheme adopted here, which predominantly describes the explicitly triggered, totally symmetric Franck-Condon dynamics. Thus, regardless of state ordering, we will miss the gradual population transfer to 1$^1$B$_{\mathrm{2u}}$, which is known to occur in the molecule after photo-excitation of 1$^1$B$_{\mathrm{3u}}$.~\cite{neuwahl1997, raytchev+2003jpca, borrego-varillas+2018as, aleotti+2021jcp} We therefore assume that our approach is valid only for a few vibrational cycles after laser excitation.

The spectrum computed for neutral pyrene (Figure~\ref{fig:abs_spec}, left panel) agrees well with experiments: Photoluminescence spectra in ethyl alcohol~\cite{Ritter2020}, absorption spectra in cyclohexane~\cite{Crawford2011}, and UV absorption spectra in acetonitrile~\cite{Jones1988} indicate \ce{P1} around 335~nm ($\sim$3.70~eV) and \ce{P2} at 272~nm ($\sim$4.56~eV).
Hence, our ALSDA result underestimates the energy of \ce{P1} and \ce{P2} by 0.28~eV and 0.34~eV, respectively, but it reproduces well their relative oscillator strengths. 
Global hybrid functionals yield excitation energies even closer to the experimental values~\cite{benkyi+2019pccp}, as expected.
From our LR-TDDFT results, we notice that the two excitations corresponding to \ce{P1} and \ce{P2} carry the majority of the oscillator strength among the first seven excited states (see Table~\ref{table:neutral+cation}).
This is expected considering that optical transitions are allowed only from the \textit{gerade} ground state, \ce{S0}, to \textit{ungerade} excited states. The transition to state \ce{S2} is parity-allowed, but is very weak as it arises from the \textit{destructive} superposition of the HOMO$-$1~$\rightarrow$~LUMO and HOMO~$\rightarrow$~LUMO+1 contributions. 

\begin{table}[h!]
\begin{tabular}{@{}ccccccc@{}}
\toprule \toprule
\textbf{$k$ in S$_k$} & \textbf{State} & \multicolumn{2}{c}{\textbf{Energy [eV]}} & \textbf{OS} & \textbf{Polarization} & \textbf{Peak label} \\ \cmidrule(lr){3-4}
  &   & \textbf{LR-TDDFT}   & \textbf{RT-TDDFT}      &    &     &   \\ \midrule
0  &  1$^\mathrm{1}$A$_\mathrm{g}$   & 0  &   -  & -     & - & - \\
\textbf{1}  & \textbf{1}$^\mathbf{1}\mathbf{B_{3u}}$ & \textbf{3.49}    & \textbf{3.42}   & \textbf{0.22}                              & $\mathbf{\hat{x}}$ &   \textbf{\ce{P1}}    \\
2  & 1$^\mathrm{1}$B$_{\mathrm{2u}}$    & 3.54      & -                      & 0                               & - &   -   \\
3   &  1$^\mathrm{1}$B$_{\mathrm{1g}}$       & 3.88    & -                        & 0                           & - &   -  \\
\textbf{4}  &  \textbf{2}$^\mathbf{1}\mathbf{B_{2u}}$    & \textbf{4.28}  & \textbf{4.22}     & \textbf{0.17}                            & $\mathbf{\hat{y}}$ &   \textbf{\ce{P2}}      \\
5    &  2$^\mathrm{1}$B$_{\mathrm{1g}}$      & 4.32  & -                          & 0                               & - &   -  \\
6   &  2$^\mathrm{1}$A$_\mathrm{g}$      & 4.46   & -                         & 0                               & - &  - \\
7  &  3$^\mathrm{1}$A$_\mathrm{g}$      & 4.77     & -                       & 0                                & - &  - \\ 
\toprule 
\textbf{$k$ in D$_k$} & \textbf{State} & \multicolumn{2}{c}{\textbf{Energy [eV]}} & \textbf{OS} & \textbf{Polarization} & \textbf{Peak label} \\ \cmidrule(lr){3-4}
  &   & \textbf{LR-TDDFT}   & \textbf{RT-TDDFT }     &    &     &   \\ \midrule
0  &  1$^\mathrm{2}$B$_{\mathrm{3g}}$   & 0   & -     & -     & - &   -  \\
1   &  1$^\mathrm{2}$B$_{\mathrm{2g}}$   & 1.06    & -         & 0   & - &  -   \\      
2  &  1$^\mathrm{2}$B$_{\mathrm{1u}}$   & 1.64    & -         & 0.02    & $\mathrm{\hat{y}}$ &  p$_1$  \\     
\textbf{3}  &  \textbf{1}$^\mathbf{2}\mathbf{A_{u}}$    & \textbf{2.13}    & \textbf{2.10}    & \textbf{0.01}   & $\mathbf{\hat{x}}$ & \textbf{\ce{p2}}    \\     
4    &  2$^\mathrm{2}$B$_{\mathrm{1u}}$     & 2.59   & -          & 0.01    & $\mathrm{\hat{y}}$ &  p$_3$  \\     
5  &  1$^\mathrm{2}$B$_{\mathrm{3g}}$     & 2.66    & -         & 0   & - &  -  \\     
6   &  1$^\mathrm{2}$A$_{\mathrm{g}}$    & 2.82   & -          & 0   & - &  - \\      
\textbf{7}   &  \textbf{2}$^\mathbf{2}\mathbf{A_{u}}$   & \textbf{3.10}      & \textbf{3.04}  & \textbf{0.22}  & $\mathbf{\hat{x}}$ &  \textbf{\ce{P1}}  \\       %
\bottomrule \bottomrule
\end{tabular}
\caption{Analysis of the ground state and of the first seven excited states of neutral pyrene (top) and its cation (bottom): symmetry, excitation energies from both LR-TDDFT (for all excitations) and RT-TDDFT (for selected excitations), oscillator strength (OS), polarization direction, and peak label are indicated. Excitations further explored in this work are highlighted in bold.}
\label{table:neutral+cation}
\end{table}

The linear absorption spectrum of the cation exhibits a few remarkable differences compared to that of the neutral molecule (see Figure~\ref{fig:abs_spec}).
While both \ce{P1} and \ce{P2} still appear as intense peaks, their energies are lower and, due to the open-shell electronic structure of the charged system, the excited states have doublet character.
LR-TDDFT calculations reveal that the composition of \ce{P1} and \ce{P2} is similar in the neutral and cationic molecule (see SI, Table~S4), although in the latter, \ce{P1} is an open-shell to vacant-shell excitation.
Below 3~eV, three weak pre-peaks appear in the spectrum of the cation (Figure~\ref{fig:abs_spec}, right panel): \ce{p1} and \ce{p3}, which are $\hat{y}$-polarized, and \ce{p2}, which is $\hat{x}$-polarized (see Table~\ref{table:neutral+cation}), in agreement with results of experiments~\cite{vala+1994,bouwman+2009aas} and calculations~\cite{vala+1994,hirata+1999jcp} reported in the literature. 
All these excitations target the SUMO (details in Table~S4) and, therefore, correspond to closed-shell to open-shell transitions. 
We note that, while spin contamination can be problematic for single-reference excited-state calculations of doublet systems, this does not apply to excitations mainly involving the open shell, but rather to closed-shell to vacant-shell transitions.~\cite{li2011jcp}

\subsection{Laser-Induced Vibronic Activity}

\begin{figure}
    \centering
    \includegraphics{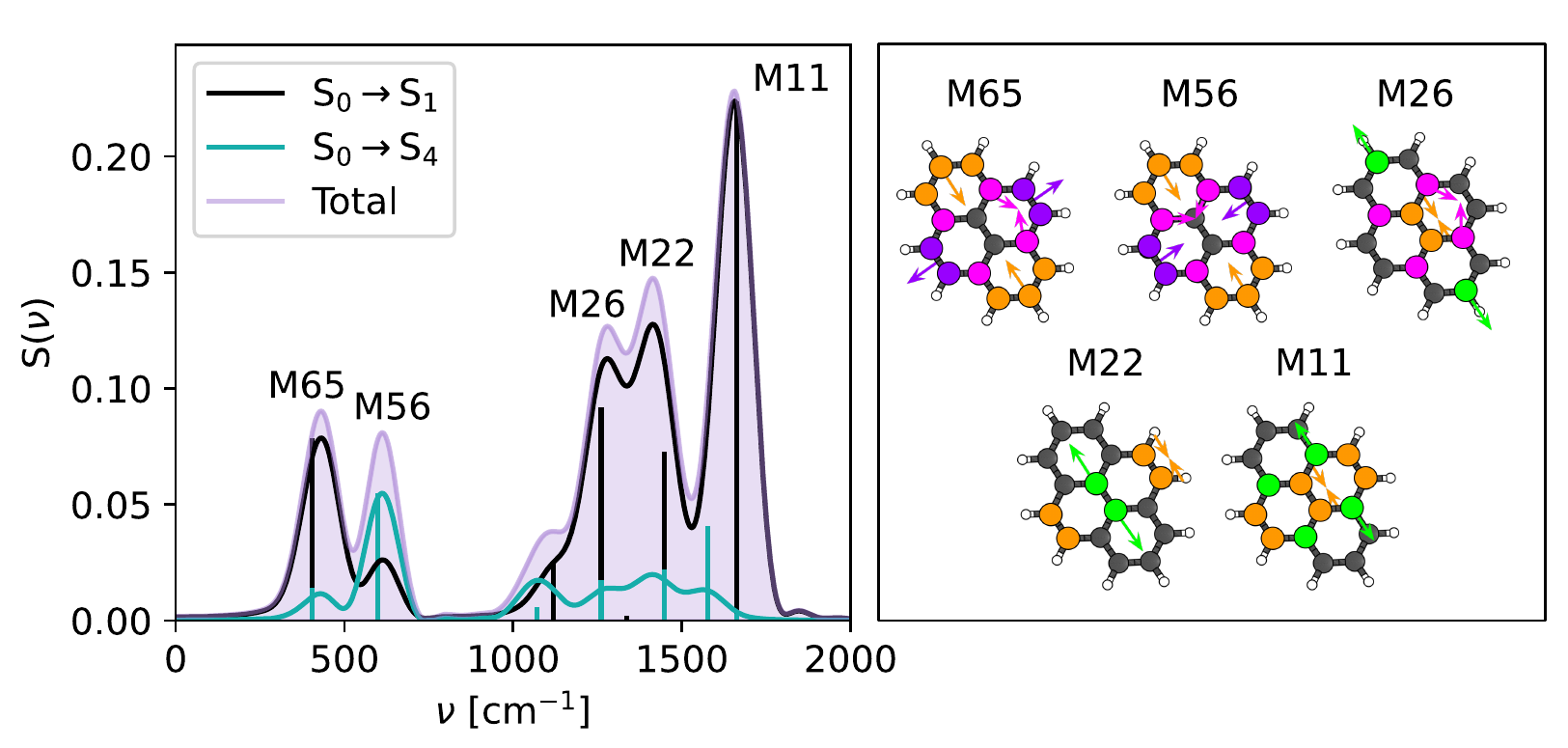}
    \caption{Left: Total KESD, $S(\nu)$, calculated for neutral pyrene upon resonant excitation of \ce{P1} (S$_0 \rightarrow$~S$_1$, black curve) and \ce{P2} (S$_0 \rightarrow$~S$_4$, green curve) as a function of the photoexcited vibrational frequencies $\nu$, and their sum indicated by the shaded purple area. Right: Normal modes excited in these dynamics marked by different colors: orange/green denotes motion along the \textit{x}-axis, purple denotes motion along the \textit{y}-axis (see coordinate system in Figure~\ref{fig:nrg}), and pink denotes off-axis motion.}
    \label{fig:Np1p2}
\end{figure}

In this section, we illustrate the results of time-dependent calculations, wherein neutral and cationic pyrene are excited by a Gaussian pulse in resonance with selected excitations (bolded rows in Table~\ref{table:neutral+cation}) and the nuclei are free to move. 
The total kinetic energy spectral density (KESD, see Eq.~\ref{eqn:power}) computed for neutral pyrene as a function of vibrational frequencies $\nu$ is displayed in Figure~\ref{fig:Np1p2}, left panel, for pulses in resonance with
\ce{P1} (black curve) and \ce{P2} (green curve). 
The vertical lines are generated by calculating the differences of the atomic positions in the geometries optimized in the GS and in the respective excited-state, and by projecting these differences onto the GS normal modes\cite{negri1994jcp}. 
In order to make the relative peak heights comparable, these projections are scaled by the corresponding normal frequencies, since the KESD is calculated from the normal velocities $\{\dot{\cal Q}_k\}$ rather than from the displacements $\{{\cal Q}_k\}$, and $\dot{\cal Q}_k\sim i\omega_k{\cal Q}_k$, \textit{i.e.}, the magnitudes of the normal displacements and velocities differ by a factor of $\omega_k$.
The results from the two approaches are in very good agreement; discrepancies likely arise from numerical differences within the codes and basis sets (since the normal modes are computed with Gaussian 16~\cite{g16} -- see Section~\ref{ssec:comput}).

From the KESD plotted in Figure~\ref{fig:Np1p2}, we notice that the laser-induced nuclear motion is harmonic regardless of whether pyrene is excited from the ground state \ce{S0} to \ce{S1} (\ce{P1}) or to \ce{S4} (\ce{P2}); the photoexcited vibrational frequencies, $\nu$, coincide with those of the normal modes of pyrene, $\nu_0$ (see Figure~S4 of the SI).
The principal vibrational motions contributing to the KESD are visualized on the right side of Figure~\ref{fig:Np1p2}, labeled by their normal mode number (see Table~S5 in the SI).
In the low-frequency region of the total KESD spectrum (Figure~\ref{fig:Np1p2}, left), maxima appear at the frequencies of the breathing modes, M65 and M56.
Interestingly, both modes are excited when the pulse is in resonance with both \ce{P1} and \ce{P2}, although the former (latter) mode is mainly activated in the S$_0 \rightarrow$~S$_1$ (S$_0 \rightarrow$~S$_4$) transition.
In the higher-frequency region of the KESD spectrum obtained by exciting \ce{P1}, peaks correspond to C-C stretching motions and to ``ring deform'' modes, which are neither breathing modes nor C-C stretching modes but have some qualities of both. 
 Mode M26 is mainly an oscillating constriction of the two outermost rings, while mode M22 features a strong stretching motion between the two central carbons; finally, mode M11 is a pure carbon-carbon stretch in the $\hat{\textit{x}}$-direction. 
All the aforementioned vibrational modes have $a_g$ symmetry, as required by Franck-Condon selection rules (see SI, Table~S5). 
The frequency range between 1000~cm$^{-1}$ and 2000~cm$^{-1}$ features only weak maxima when pyrene is excited in resonance with \ce{P2}.
The stronger vibrational response of pyrene to pumping at \ce{P1} than at \ce{P2} is a consequence of the transition dipole of \ce{P1} being oriented in the $\hat{\textit{x}}$ direction,which, in turn, is a consequence of the increasing number of orbital nodal planes along $\hat{\textit{x}}$ associated with the HOMO-LUMO transition making up \ce{P1}. This entails strong excitation-induced charge and potential gradients in the $\hat{\textit{x}}$ direction, giving rise to a force field that can effectively stimulate stretching motions of parallel C-C bonds (see SI, Figure~S6). It corresponds to the potential gradients driving the wavepacket away from the Franck-Condon region of the excited-state potential-energy surface. The direction of the induced motion -- bond expansion or compression -- is determined by the positions of charge accumulation and depletion, which attract or repel the positively charged nuclei, respectively, pulling or pushing them out of their equilibrium position. Gradients in the perpendicular $\hat{\textit{y}}$ direction are also present, but significantly smaller in magnitude. In turn, \ce{P2} with its $\hat{\textit{y}}$-oriented transition dipole moment leads to forces mainly along $\hat{\textit{y}}$; however, since no bonds are oriented along $\hat{\textit{y}}$, these excitation-induced charge gradients cannot drive bond vibrations as effectively (see SI, Figure~S7), and thus primarily activate the collective breathing modes.
Both excitations also result in forced vibrations of anti-symmetric modes, which oscillate at the frequency $\nu$ of the pump (see SI, Figure~S4). They are driven by and thus have the same symmetry of the coherently oscillating electron density, which is equal to the transition density of the excitation. 

\begin{figure}[H]
    \centering
    \includegraphics[width=0.5\textwidth]{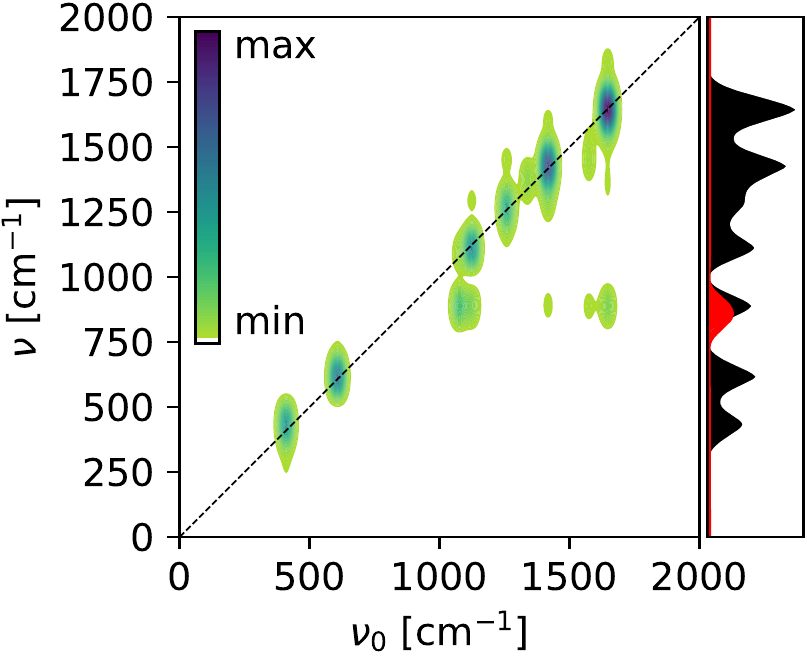}
    \caption{Left: Kinetic energy spectral density, $S_k(\nu)$, of the pyrene cation pumped at \ce{P1} as a function of the real-time vibrational frequencies, $\nu$, and of the normal frequencies $\nu_{0}$. An artificial Gaussian broadening with $\sigma=20~\mathrm{cm}^{-1}$ is applied along the $\nu_0$ axis. Right: Total kinetic energy spectral density $S(\nu)$, with the Gaussian-filtered Fourier transform of the electric dipole moment envelope displayed in red.
    }
    \label{fig:cation_p1_power}
\end{figure}

The photo-induced nuclear dynamics of the pyrene cation pumped at \ce{P1} is characterized by a prominent anharmonic feature. 
To analyze this characteristic in detail, we plot the KESD, $S_k(\nu)$, as a function of both the real-time photo-induced frequencies, $\nu$, and the frequencies of the normal modes, $\nu_{0}$ (see Figure \ref{fig:cation_p1_power}, left panel).
The dominant contributions to the KESD arise from vibrational modes with $a_g$ symmetry, and are located along the diagonal of the KESD, hinting at harmonic motion.
However, at $\nu \approx$~900~cm$^{-1}$, off-diagonal signals appear and give rise to a distinct peak in the total KESD (see right panel of Figure \ref{fig:cation_p1_power}).
The analysis of the coupled electron-vibrational dynamics reveals that at this frequency the nuclear motion is correlated with the beating of the induced dipole moment (red peak in Figure \ref{fig:cation_p1_power}, right panel).

\begin{figure}[h!]
    \centering
    \includegraphics[width=0.5\textwidth]{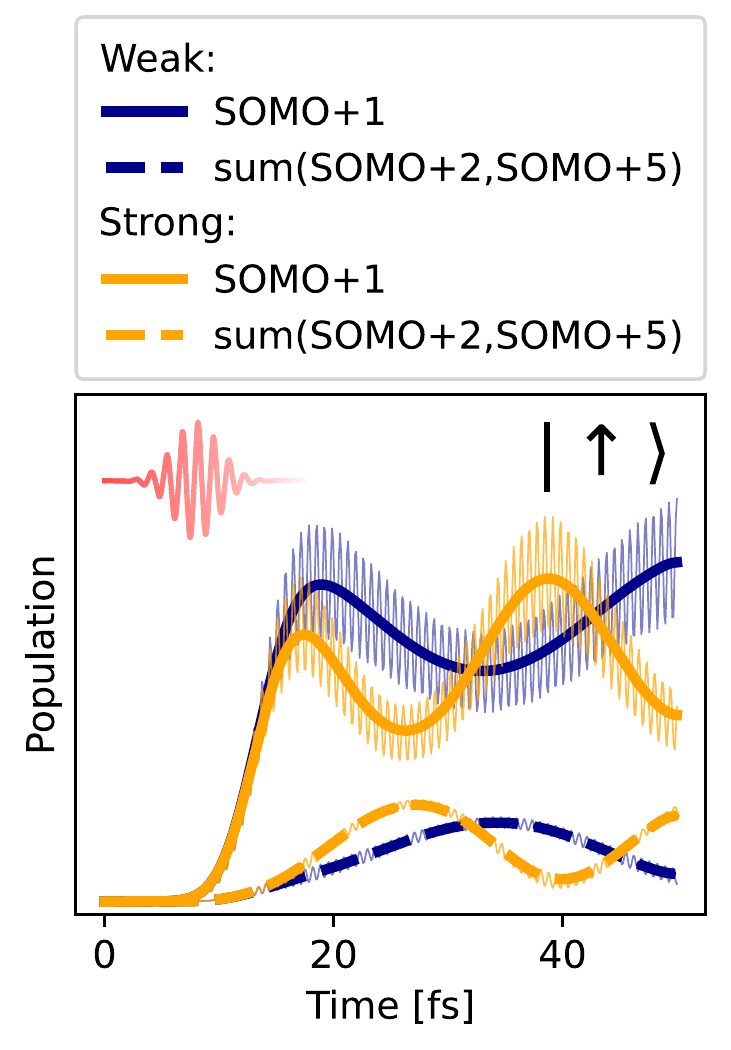}
    \caption{Time-dependent intensity-normalized electronic populations of the virtual spin-up orbitals following laser-induced \ce{D0}$\rightarrow$\ce{D7} excitation of the pyrene cation. Blue and orange curves correspond to laser intensities of $1\times 10^{10}$~W/cm$^2$ (\textit{weak}) and $4\times 10^{10}$~W/cm$^2$ (\textit{strong}), respectively. The faint oscillatory lines are the raw projections, while the solid bold lines show the data filtered with a Gaussian curve.
    }
    \label{fig:pops}
\end{figure}

To determine the origin of the beating that is present in the photo-excited system even in the absence of vibronic dynamics, we calculate the TD spin-up populations of the GS orbitals according to Eq.~\eqref{eqn:orbitalProjection} in the fixed-nuclei scenario (Figure~\ref{fig:pops}). The virtual orbitals mainly contributing to the \ce{D0}$\rightarrow$\ce{D7} excitation (SOMO+1 and SUMO, see Table~S4) periodically exchange electronic population with higher unoccupied orbitals, including those which do not play a significant role in any linear-regime excitation below 4~eV, \textit{e.g.}, the SUMO+5. Given the absence of time-dependent external fields following the pump pulse, these Rabi oscillations are driven by dynamical electron-electron interactions. Furthermore, the population oscillation amplitudes depend non-linearly on electric field intensity, indicating that this is a higher-order phenomenon.
This interpretation is clearly supported by the results shown in Figure~\ref{fig:pops}. As the displayed electronic populations are intensity-normalized, the blue (weak field intensity) and orange (strong field intensity) curves would have had the same amplitude in case of linear dependency between amplitude and intensity. This not being the case is a signature of non-linear dependency.

\begin{figure}
    \centering
    \includegraphics[width=0.5\textwidth]{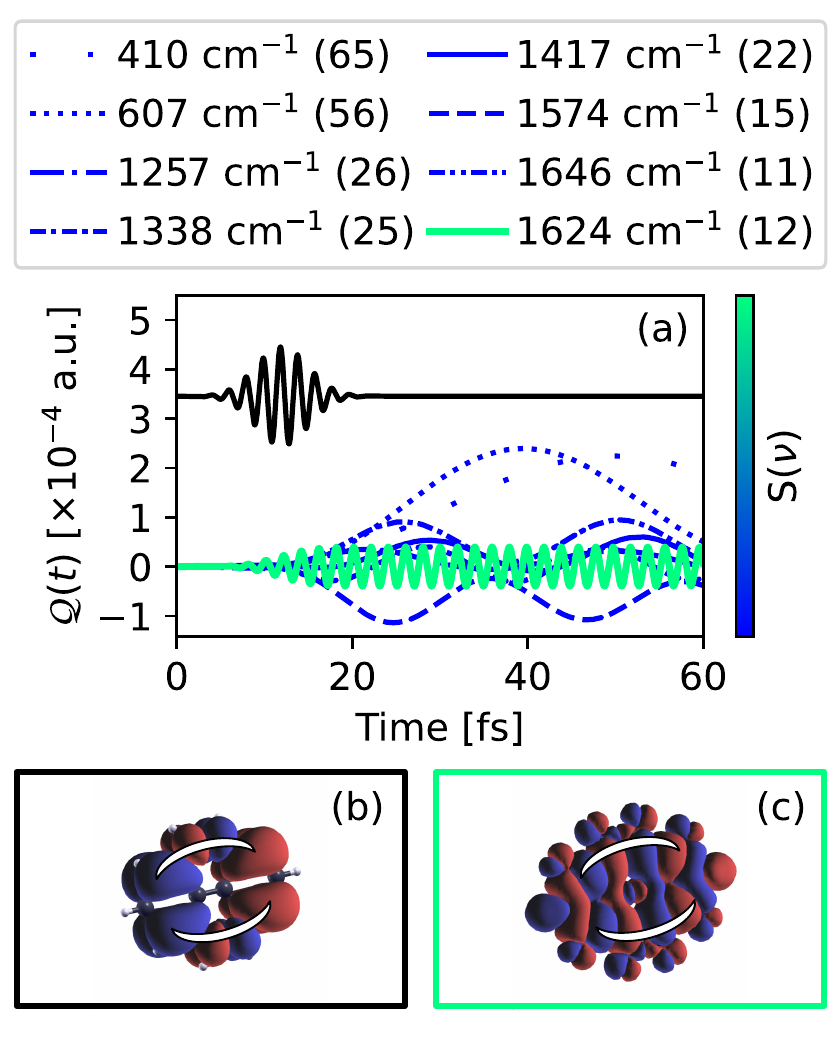}
    \caption{Pyrene cation excited at the $\hat{x}$-polarized pre-peak, \ce{p2}. (a) Time evolution of the normal displacements ${\cal Q}(t)$ along the displayed selected modes. The color bar is associated with the KESD, $S(\nu)$: the seven most relevant symmetric modes are in blue and the anti-symmetric mode (the most intense overall) in green. The black curve shows the applied Gaussian laser pulse with energy $\hbar \omega =$~2.1~eV. (b) Transition density of the excitation at 2.1~eV, and (c) approximate charge flux due the vibration of mode 12. The white arcs in these plots highlight similarities in the densities and their nodal planes.
    }
    \label{fig:cation_prepeak}
\end{figure}

Finally, we examine the ultrafast vibrational response of the pyrene cation pumped at 2.1~eV in resonance with one of its pre-peaks, \ce{p2} (see Figure \ref{fig:cation_prepeak}). 
The resulting nuclear dynamics are dominated by seven symmetric modes with $a_g$ symmetry (modes 65, 56, 26, 25, 22, 15, and 11 -- see Table~S5), as well as by one anti-symmetric mode with $b_{3u}$ symmetry (mode 12) which yields the largest overall contribution to the total KESD.
Thus, the normal velocity associated with this mode is much higher than that of the harmonically oscillating ones, whereas the opposite is true for the normal displacements. 
The anti-symmetric mode does not depart far from the equilibrium position, and it therefore has large kinetic energy but low potential energy.

The participation of both symmetric and anti-symmetric modes to the ultrafast dynamics of the pyrene cation can be rationalized considering a two-level system as a minimal model for the resonantly driven electronic system. In the Ehrenfest scheme, the forces on the nuclei are the Coulomb repulsion from the other nuclei and the attraction of the averaged electronic cloud. In the beginning, these forces cancel out as the molecule rests in equilibrium. Due to the ultrashort nature of the external electric field, there is a brief time window after the pulse in which the nuclei are still in their equilibrium positions, but the electrons are in a non-stationary superposition state. During this time, the two-level approximation with fixed electronic energy levels is valid, and in this framework the induced electron density is given by\cite{Krumland2020}
\begin{equation}
\begin{split}
    \delta \rho(\mathbf{r},t) &= \rho(\mathbf{r},t) - \rho_{\mathrm{GS}}(\mathbf{r})\\
    &= |c_{\mathrm{ES}}|^2 [\rho_{\mathrm{ES}}(\mathbf{r})-\rho_{\mathrm{GS}}(\mathbf{r})] + 2\Re [c^*_{\mathrm{GS}}c_{\mathrm{ES}}e^{-i(E_{\mathrm{ES}}-E_{\mathrm{GS}})t}\rho_{\mathrm{GS}\rightarrow\mathrm{ES}}(\mathbf{r})],
\end{split}
    \label{eqn:dens2}
\end{equation}
where $\rho_\mathrm{GS}$ and $\rho_\mathrm{ES}$ are the stationary electron densities of the GS and of the vertically excited state, respectively, and $\rho_{\mathrm{GS}\rightarrow\mathrm{ES}}$ is the transition density between them. The wavefunction expansion coefficients in the eigenstate basis, $c_\mathrm{GS}$ and $c_\mathrm{ES}$, are expressed in the interaction picture, where they are time-independent after the pulse. 

The two terms on the right-hand-side of Eq.~\eqref{eqn:dens2} give rise to forces that drive the nuclei away from their equilibrium positions. The first term -- depending on the density difference and already mentioned above in the discussion of the difference between the vibronic couplings of \ce{P1} and \ce{P2} -- corresponds to a sudden displacement of the energetic minimum of the averaged potential energy surface, on which the dynamics occur in the adopted Ehrenfest scheme. The nuclei, still in their original equilibrium position, start to oscillate around the new minimum at their fundamental frequencies, which may be slightly altered due to the mixing in of excited-state potential energy surfaces with different curvatures.\cite{hernandez2019jpca}
$\rho_\mathrm{GS}$ and $\rho_\mathrm{ES}$ are calculated from the square moduli of the many-electron wave functions and are thus totally symmetric. Consequently, $\rho_\mathrm{ES}-\rho_\mathrm{GS}$ and the corresponding electrostatic forces on the nuclei are also totally symmetric, and thus activate only normal modes of $a_g$ symmetry, in accordance with the Franck-Condon selection rules. The energy of these modes is proportional to the excited state population $|c_\mathrm{ES}|^2$, which in turn depends on the intensity $|E_0|^2$ of the driving field. 

The second term in Eq.~\eqref{eqn:dens2} is time-dependent, oscillating at the transition frequency $E_{\mathrm{ES}}-E_{\mathrm{GS}}$. This is a rapidly oscillating force associated with the electronic coherence, which drives the nuclear motion at a much higher frequency than its resonances in the infrared. This force is often weak and therefore negligible. For this reason, excited-state molecular dynamics simulations initially prepare the system in a stationary excited state, implicitly neglecting such terms. Here, however, the forced motion has an amplitude comparable to the fundamental vibrations, and a much larger associated kinetic energy. The transition density $\rho_{\mathrm{GS}\rightarrow\mathrm{ES}}$, which gives the spatial dependence of the oscillating density, is proportional to the product of the two states. Its symmetry, given by the direct product of the corresponding irreducible representations, is the same as that of the transition dipole moment and, in our case, also of the polarization direction of the exciting laser field. As a consequence, the operator describing the force exerted on the nuclei due to $\rho_{\mathrm{GS}\rightarrow\mathrm{ES}}$ is of $b_{3u}$ symmetry, and only affects corresponding normal modes. This force has a transient nature, as it is related to the presence of electronic coherence, which is damped over a couple of tens of fs due to electronic dephasing. In our calculations, incoherent processes inducing this decoherence are not included: hence, the corresponding density oscillations persist indefinitely. The term has a linear dependency on the field intensity: $c_\mathrm{ES} \propto |E_0|$. Therefore, the relative strength of the two types of nuclear motion depends on the field intensity, and one could tune the electric field amplitude in order to equalize the power in the symmetric and anti-symmetric modes, without qualitatively affecting the features of the power spectrum.

We emphasize that the activation of non-totally-symmetric vibrations is not a Herzberg-Teller effect, which can also be associated with such modes. Similar to the case of vibrationally-induced transitions after the laser pulse mentioned above, Herzberg-Teller vibronic coupling requires a finite initial normal mode displacement, which would be given in a fully quantum-mechanical picture, as the ground-state nuclear wavefunction has an extension around the minimum. With purely classical nuclei at 0K, however, the system is initially localized at ${\cal Q}=0$ for all modes. Therefore, such post-Franck Condon vibronic couplings are prohibited. 
The effect described here is different; it is a transient forced motion that requires the system to be in a non-stationary electronic superposition state, in contrast to Franck-Condon and Herzberg-Teller couplings which are associated with free vibrations of the normal modes.

The driven mode has strong infrared activity due to coupling to intramolecular charge fluxes, which give rise to a strong oscillation of the total dipole moment.\cite{torii+1999jpca} Here, we notice the reverse effect: $\rho_{\mathrm{GS}\rightarrow\mathrm{ES}}$ (Figure \ref{fig:cation_prepeak}b) shares similarities with the charge flux, $\rho_\leftrightarrow$, associated with the vibration, and thus drives it. This flux can be estimated by subtracting the equilibrium GS density from the GS density after deformation of the molecular geometry along the direction of the normal mode (Figure \ref{fig:cation_prepeak}c). While the transition density and the charge flux are quite distinct at first glance, a closer inspection shows that the nodal planes of $\rho_{\mathrm{GS}\rightarrow\mathrm{ES}}$ and $\rho_\leftrightarrow$ coincide along the highlighted paths, such that the resulting forces on the nuclei are qualitatively similar in those regions. Thus, the rapidly oscillating transition density gives rise to an equally rapid vibration of the normal mode, far away from its resonance.

\subsection{Transient Absorption Spectra}

\begin{figure}[h!]
    \centering
    \includegraphics{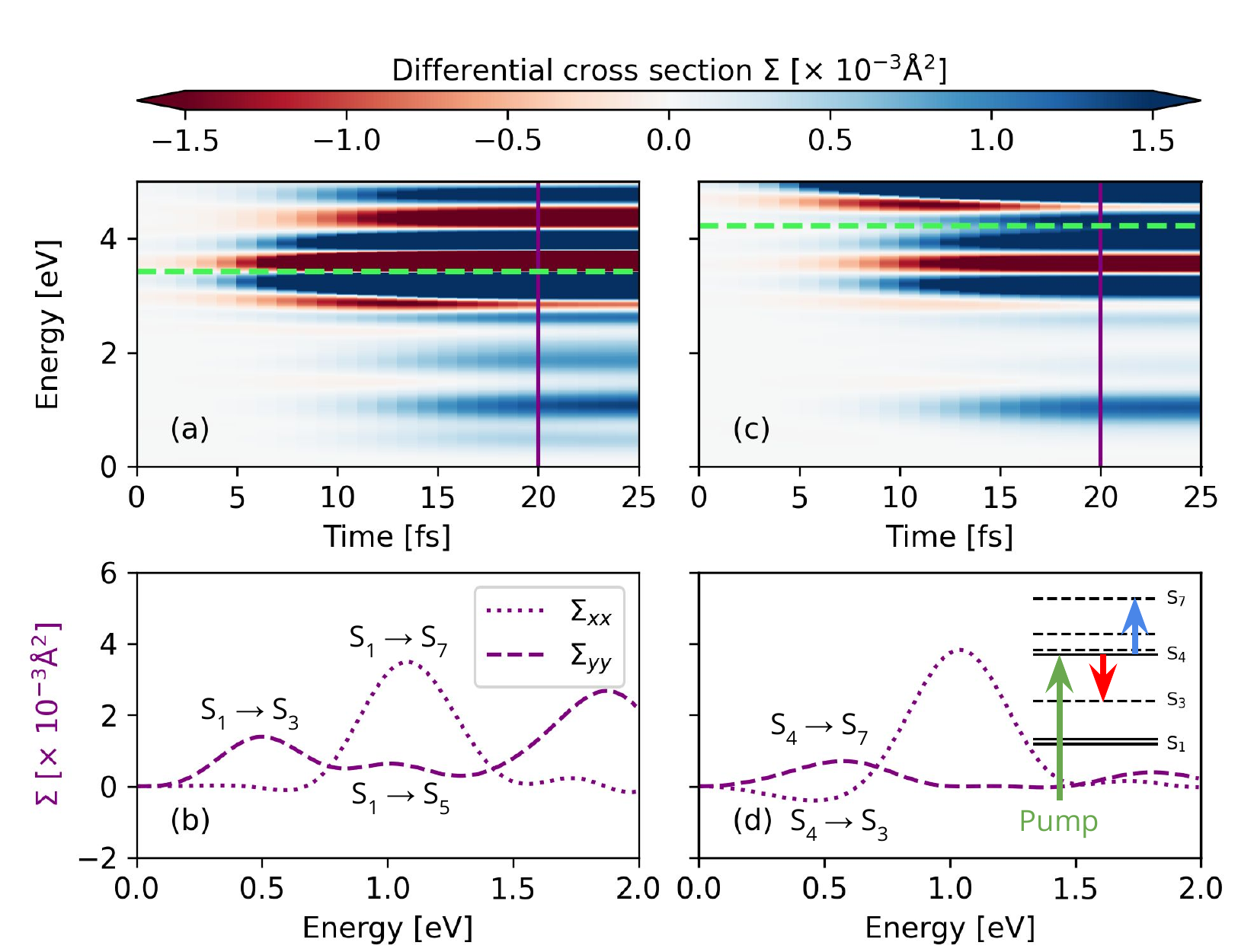}
    \caption{Stationary-ion TAS for neutral pyrene pumped at (a) \ce{P1} and (c) \ce{P2}. The purple line at 20 fs indicates the time at which the differential cross section $\Sigma$ is plotted in panels (b) and (d), where the $xx$- and $yy$-components of $\Sigma$ (which is a tensor) are marked by dotted and dashed lines, respectively. The excited-state transitions giving rise to the peaks are indicated, according to the scheme in the inset of panel (d), where solid (dashed) lines represent $u$ ($g$) states.}
    \label{fig:neutral_tas_stationary}
\end{figure}

The results collected so far equip us to analyze the transient absorption spectra (TAS) of pyrene and its cation. 
We examine separately the TAS computed with stationary ions and the TAS obtained by including the nuclear motion. In this way, we can pinpoint the electronic and vibrational contributions to the differential cross sections. 

We start from the analysis of the TAS calculated for neutral pyrene pumped at \ce{P1} keeping the nuclei stationary (Figure \ref{fig:neutral_tas_stationary}a). 
From the differential cross section relative to the spectrum at $t_0=0$~fs, we notice that after the introduction of the electric field, the pumped peak \ce{P1} red-shifts as a known artifact of the adiabatic approximation.\cite{fuks2015prl} 
In the low-energy region, a few pre-peaks emerge. Symmetry considerations are useful in relating these features to excited-state absorption processes: the transition \ce{S1}$\rightarrow$ S$_\text{k}$ is symmetry-allowed and polarized along $\hat{x}$ or $\hat{y}$ if the direct product of the symmetries of \ce{S1} and S$_\text{k}$ yields B$_{3u}$ or B$_{2u}$, respectively. This information and the difference of the excited-state energies obtained from the linear-response calculation are used to analyze the polarization-resolved differential cross sections plotted in Figure~\ref{fig:neutral_tas_stationary}b. This comparison enables an unambiguous assignment of the first pre-peaks: Due to the coupling to the states \ce{S3} and \ce{S7}, the population of the first excited state, \ce{S1}, is responsible for these near-IR absorption features. When neutral pyrene is pumped at \ce{P2} (Figure \ref{fig:neutral_tas_stationary}c), again, several pre-peaks emerge at low energy: The strongest is at approximately 1~eV, and is due to absorption from \ce{S4} to a manifold of higher excited states. Approaching the continuum, these states become increasingly dense, thereby complicating the assignment. Between 0.4~eV and 0.5~eV, contributions from an $\hat{\textit{x}}$-polarized negative peak (\ce{S4}$\rightarrow$\ce{S3}, corresponding to stimulated emission) and a $\hat{\textit{y}}$-polarized positive peak (\ce{S4}$\rightarrow$\ce{S7}) effectively cancel each other, leading to a vanishing differential cross section in the polarization-averaged TAS (see Figure ~\ref{fig:neutral_tas_stationary}d). Hence, the excited states \ce{S3} and \ce{S7} shape the non-linear absorption processes following UV excitations in resonance with both \ce{P1} and \ce{P2}. 

\begin{figure}[t]
    \centering
    \includegraphics[width=\textwidth]{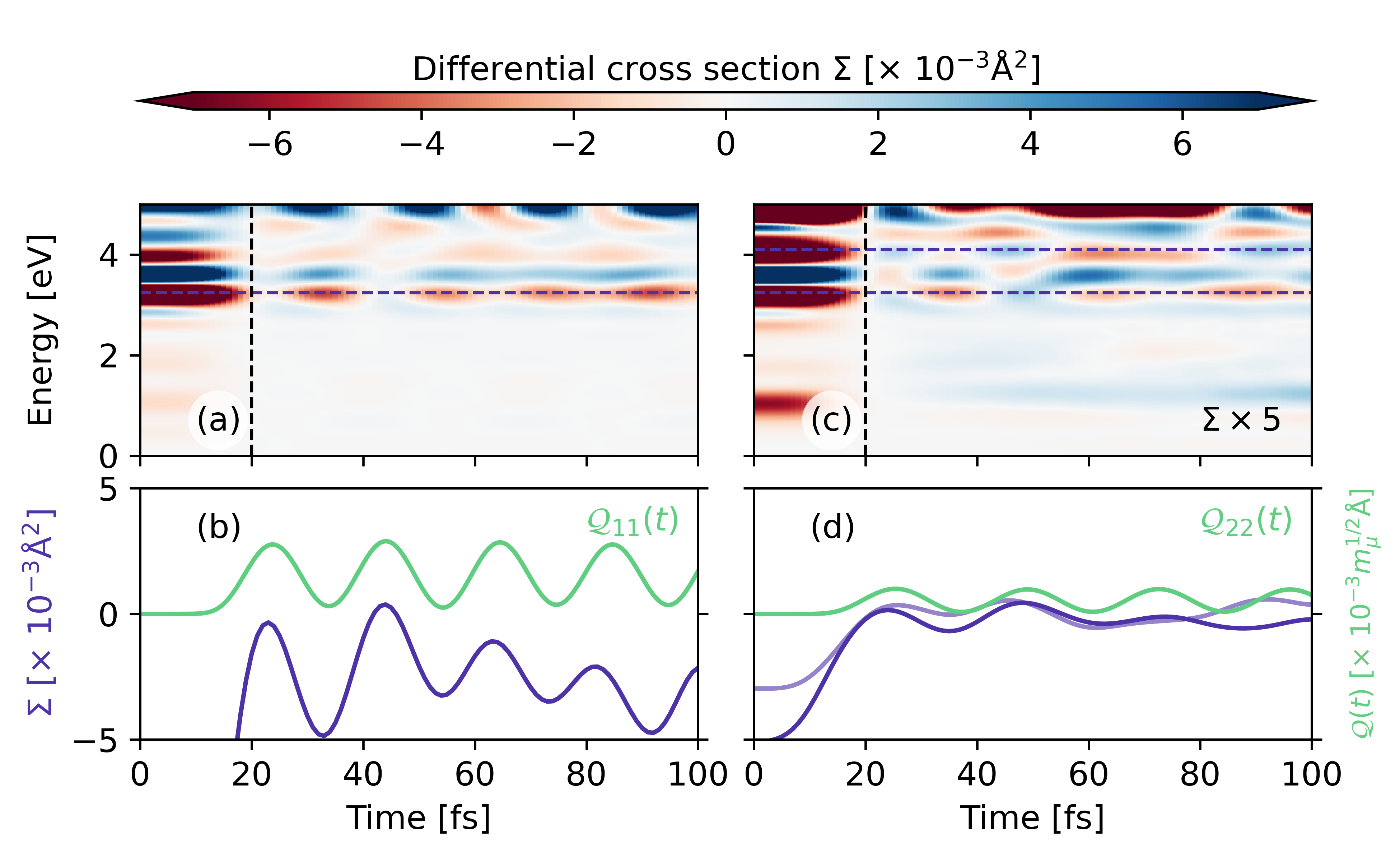}
    \caption{TAS of neutral pyrene with moving ions where the differential cross section $\Sigma$ is plotted relative to $t_0$=20~fs pumped at (a) \ce{P1} and (c) \ce{P2}. The differential cross section $\Sigma$ at the energies marked by the dashed horizontal lines in panels (a) and (c) is shown in purple in panels (b) and (d), respectively. These energy cuts at 3.24 eV and 4.10 eV are chosen close to the pumping energies. The green lines are the displacements ${\cal Q}(t)$ of mode 11 in (b) and mode 22 in (d), which are highly correlated to the TAS cuts.
    }
    \label{fig:fig10}
\end{figure}

We now extend our analysis of neutral pyrene in order to examine the influence of nuclear motion on the TAS (Figure~\ref{fig:fig10}). When the molecule is pumped at \ce{P1} (Figure~\ref{fig:fig10}a), all features in the TAS are modulated by the normal vibrational mode M11 with frequency $\nu_0 =$~1647~cm$^{-1}$. This behavior is evident by inspecting Figure~\ref{fig:fig10}b: The differential cross section at the energy indicated by the horizontal dashed line (purple curve) and the mode displacement according to Eq.~\eqref{eqn:normalTrafo} [${\cal Q}_{11}(t)$, green curve] have nearly coincident maxima and minima. This is unsurprising, since the KESD in Figure~\ref{fig:Np1p2} shows that M11 has the most power. 
Next, we examine the TAS of the neutral molecule excited in resonance with the energy of \ce{P2}. The magnification factor 5 adopted in Figure~\ref{fig:fig10}c indicates that the values of the differential cross section relative to 20~fs are small compared to the TAS in Figure~\ref{fig:fig10}a. This is reasonable considering that the vibrational motion initiated by exciting \ce{P2} is much weaker with respect to the motion for pumping \ce{P1}, as evidenced again by the KESD in Figure~\ref{fig:Np1p2}. 
Although M22 does not provide the strongest contribution to the KESD, it is strongly correlated to the TAS cuts, as seen in Figure~\ref{fig:fig10}d. These results agree well with a previous experimental study, wherein a heterodyne transient grating spectrum of pyrene in ethanol revealed an intensity beating with frequency 1413~cm$^{-1}$\cite{Picchiotti2019}. Notice that our calculations predict \textcolor{black}{M22} with a frequency of 1417~cm$^{-1}$.

\begin{figure}[t]
    \centering
    \includegraphics{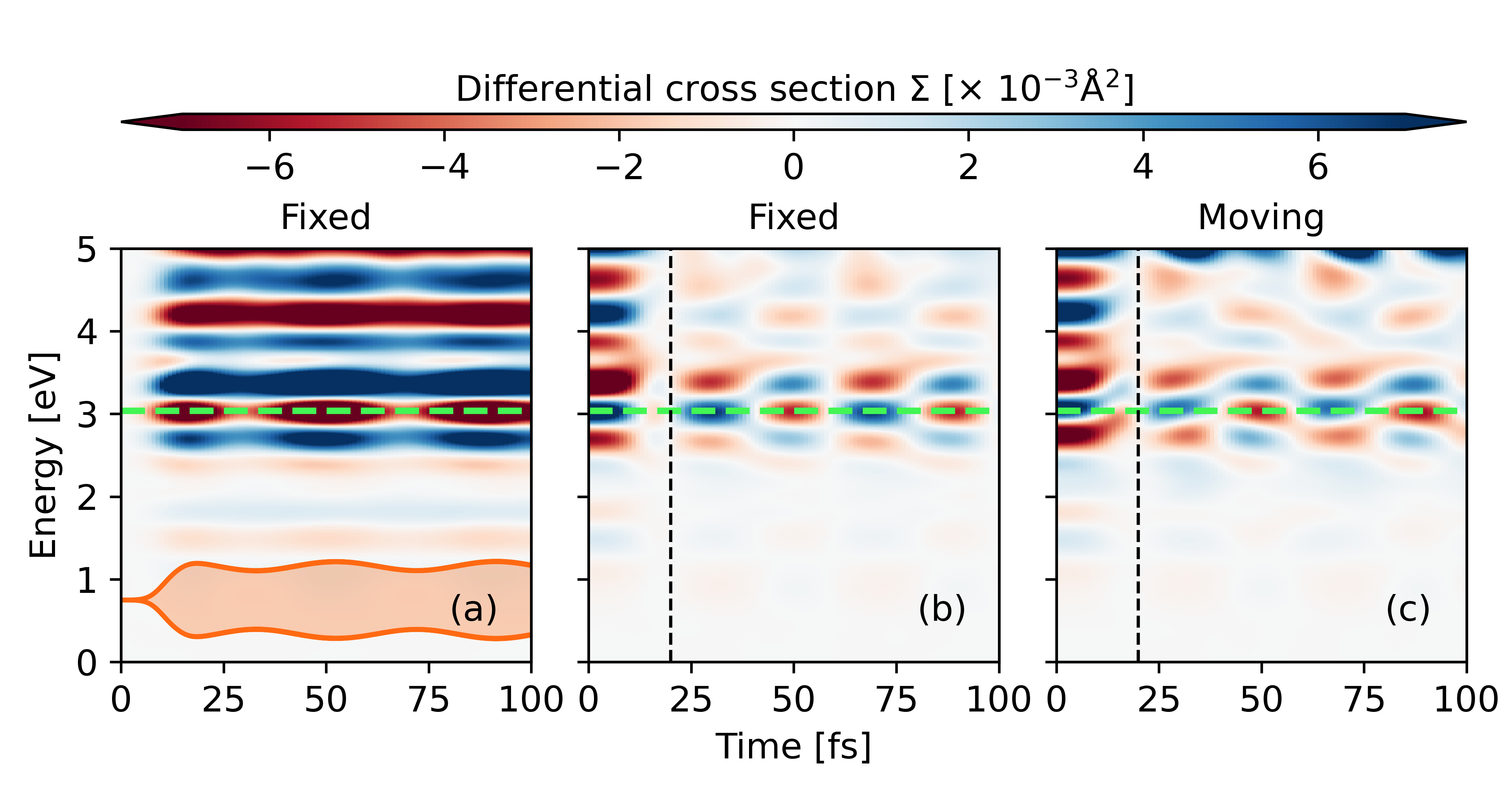}
    \caption{TAS of cationic pyrene pumped at \ce{P1} with the differential cross section $\Sigma$ calculated for (a) stationary-ions relative to 0 fs, (b) stationary-ions relative to 20 fs, and (c) moving-ions relative to 20 fs. The excitation energy of 3.04~eV is marked with a green dashed line. The orange curve in panel (a) is the $\hat{x}$-polarized dipole moment envelope (the vertical scale is arbitrary).}
    \label{fig:cation_tas_all}
\end{figure}

Analogous TAS analysis is conducted for the cation, which exhibits very different behavior compared to neutral pyrene (see Figure~\ref{fig:cation_tas_all}). When pumping at \ce{P1} (green dashed line) and holding the ions stationary, both main peaks broaden, as shown by the blue-red-blue pattern in Figure~\ref{fig:cation_tas_all}a. Simultaneously, the cross section of the three pre-peaks decreases. Notably, there is beating in the differential cross section over time, which reflects a beating in the $\hat{x}$-component of the dipole moment envelope (orange inset in Figure \ref{fig:cation_tas_all}a). 
Even when the ions are permitted to move, this electronic effect remains responsible for most of the  modulation of the TAS (Figure~\ref{fig:cation_tas_all}c).
Interestingly, the same effect also gives rise to the anharmonic vibrational response of the system (see Figure~\ref{fig:cation_p1_power}). 
Figure~\ref{fig:cation_tas_all}b (stationary-ion TAS relative to 20~fs) is included to facilitate comparison between the panels for fixed- and moving-ion TAS results. Due to the similarity of the plots in panels (b) and (c), we can confirm that this electronic beating overpowers any normal mode TAS contributions and is the driving force in all instances.


\section{Summary and Conclusions}

We have presented an extensive first-principles study on the sub-picosecond electronic and vibrational dynamics of the pyrene molecule and its cation excited by ultrafast, coherent UV pulses. 
From the analysis of the electronic structure of the systems, we gained insight into their allowed optical transitions, thereby identifying target states for resonant, time-dependent external fields. 
In parallel, we examined the vibrational activity of pyrene, by characterizing the symmetry and the intensity of its normal modes.
In this way, we could disentangle the effects related to laser-induced electronic and vibrational dynamics represented in terms of transient absorption spectra and kinetic energy spectral density.

We investigated neutral pyrene excited in resonance with the two absorption maxima in the low-energy region of its UV spectrum. 
When the molecule is pumped at the energy of its first excitation, the high-frequency C-C stretching modes are mostly excited, transferring a large amount of kinetic energy to the molecule. 
When, instead, the pulse has the frequency of the second bright excitation, which has perpendicular polarization with respect to the first one, pyrene gains less kinetic energy. 
In this case, the low frequency breathing motion is favoured and the C-C stretching modes have comparable, relatively weak contributions. 
We understand these results in terms of polarization of the pump pulse and its relation to the C-C bonds in the molecule. 

The vibronic response of the cation is characterized by driven
oscillations resulting from different manifestations of quantum interference between multiple states.
We propose that some of these effects stem from the non-linear response of the system, which leads to persistent Rabi oscillations between KS orbitals, driven by electron-electron interactions.
We additionally investigated the vibronic response of the cation to pulses in resonance with its lowest-energy excitation, which has very weak oscillator strength and appears in the visible region, below the absorption onset of the neutral molecule. 
The first-principles results of these dynamics were interpreted with the aid of an auxiliary two-level model, which helped clarify that symmetric vibrational modes are predominantly excited. 
It is worth noting that, due to their complete symmetry, these modes are never infrared active. 
This implies that infrared emission observed for pyrene and its cation in the interstellar medium~\cite{baba2009vibrational,tielens2008interstellar} occurs only within or beyond nanosecond timescales when electronic excitation energy is completely internally converted into vibrational energy (heating), or though post-Franck-Condon vibronic couplings between excited states.
Other emission mechanisms driven by the anharmonicities of the nuclear motion are also foreseeable based on the presented results.
While dedicated research is certainly needed to disclose and understand these effects, our analysis provides an essential starting point in this regard. 

In conclusion, the results of our comprehensive study contribute to a deeper understanding of the photo-response of pyrene to coherent, ultrafast perturbations. 
Our analysis reveals a non-trivial interplay between laser-driven electronic and nuclear motion in this molecule, showing that anharmonicities are pronounced in the ionized structure. 
Given the structural similarities among PAHs, our findings offer valuable hints to interpret general features of laser-induced electronic and vibronic dynamics in this class of molecules, and represent the basis for future studies on such systems, including their functionalized counterparts and their combination with inorganic substrates.

\section*{Acknowledgments}
This work was funded by the Deutsche Forschungsgemeinschaft (DFG, German Research Foundation) - project number 182087777 - SFB 951, by the German Federal Ministry of Education and Research (Professorinnenprogramm III), and by the State of Lower Saxony (Professorinnen für Niedersachsen). K.R.H. acknowledges the support of the Humboldt Internship Program and the University of Ottawa Co-operative Education Program, and thanks the Humboldt-Universit\"at zu Berlin (Santander Scholarship) and the University of Ottawa Centre for Research Opportunities for their generous funding. 
Computational resources were provided by the North-German Supercomputing Alliance (HLRN), project bep00076.

\section*{Data Availability}
All data reported in this work are available free of charge from Zenodo at the following DOI: 10.5281/zenodo.5486137.

\begin{suppinfo}
We provide the details about the structural, electronic, and vibrational properties of pyrene and its cation, as well as additional information about electronic and vibronic dynamics.

\end{suppinfo}

\bibliographystyle{achemso}
\providecommand{\latin}[1]{#1}
\makeatletter
\providecommand{\doi}
  {\begingroup\let\do\@makeother\dospecials
  \catcode`\{=1 \catcode`\}=2 \doi@aux}
\providecommand{\doi@aux}[1]{\endgroup\texttt{#1}}
\makeatother
\providecommand*\mcitethebibliography{\thebibliography}
\csname @ifundefined\endcsname{endmcitethebibliography}
  {\let\endmcitethebibliography\endthebibliography}{}

\end{document}